\documentclass[runningheads]{llncs}
\usepackage[T1]{fontenc}
\usepackage{cancel}

\usepackage{graphicx}
\usepackage{array}
\usepackage{longtable} 

\usepackage{tikz}
\usetikzlibrary{positioning,arrows.meta,fit,backgrounds}

\usepackage[utf8]{inputenc} 
\usepackage{hyperref}
\hypersetup{
  colorlinks=true,
  urlcolor=blue,
  linkcolor=black,
  citecolor=black
}
\begin{document}

\begin{titlepage}
\centering
\vspace*{1em}

\includegraphics[width=0.70\textwidth]{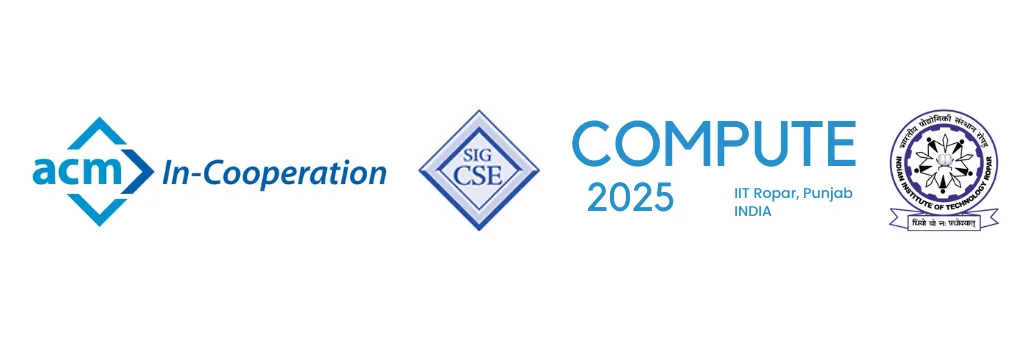}\par

\vspace{1.2em}
{\LARGE\bfseries COMPUTE 2025 -- Best Practices in Computer Science Education\par}

\vspace{0.8em}
Proceedings of the Best Practices track at the
\href{https://isigcse.acm.org/compute/2025/}{ACM COMPUTE Conference 2025}\par
Indian Institute of Technology Ropar (IIT Ropar), India\par
Conference dates: December 7--9, 2025\par
Editors: Ritwik Murali and Mrityunjay Kumar\par

\vspace{1.2em}
\hrule
\vspace{1em}

\begin{minipage}{0.92\textwidth}
This proceedings record contains the Best Practices track index and the track-level
summary report from the COMPUTE 2025 Best Practices Ideas Session. The Best Practices
sessions provide an opportunity to share teaching practices in computer science at the
higher education level in a setting that is less formal than a paper presentation.
\end{minipage}

\vspace{1.2em}
{\large\bfseries Index\par}
\vspace{0.8em}

\renewcommand{\arraystretch}{1.25}
\begin{tabular}{@{}p{0.78\textwidth}p{0.20\textwidth}@{}}
\textbf{\href{https://arxiv.org/abs/2512.03492}{Functional Python Programming in Introductory Computer Science Courses}} \\
\textit{Rajshekhar Sunderraman (Georgia State University)} & \\
\vspace{1em}

\textbf{\href{https://arxiv.org/abs/2512.03501}{SocraticAI: Transforming LLMs into Guided CS Tutors Through Scaffolded Interaction}} \\
\textit{Karthik Sunil; Aalok Thakkar (Ashoka University)} & \\
\vspace{1em}

\textbf{\href{https://arxiv.org/abs/2512.03507}{Ancient Algorithms for Modern Curriculum}} \\
\textit{Aalok Thakkar (Ashoka University)} & \\

\end{tabular}

\vfill
\end{titlepage}

%

\title{COMPUTE 2025 -- Best Practices in Computer Science Education: Ideas Session Report}

\titlerunning{COMPUTE 2025 Best Practices Proceedings}

\author{ Ritwik Murali\inst{1}\orcidID{0000-0002-1269-2257} \and
        Mrityunjay Kumar\inst{2,3}\orcidID{0000-0003-2819-759X}}

\institute{
Department of Computer Science and Engineering, Amrita School of Computing, Coimbatore, Amrita Vishwa Vidyapeetham, India \\ 
\email{m\_ritwik@cb.amrita.edu} 
\and 
Birla Institute of Technology and Science, Pilani, Rajasthan 333031, India \\ \email{mrityunjay.kumar@pilani.bits-pilani.ac.in} 
\and
International Institute of Information Technology, Hyderabad 500032, India \\
\email{mrityunjay.k@research.iiit.ac.in}
}

\maketitle              
\begin{abstract}

COMPUTE Best Practices sessions provide an opportunity to share teaching best practices in computer science in higher education, in a setting that is less formal than a paper presentation. This report summarizes the activities in the two Best Practices sessions at COMPUTE 2025 and lists the ideas shared by the attendees. Three areas were explored: (1) class activities that support engagement and learning in programming classes, (2) the design of take-home programming assignments that are resistant to AI misuse, and (3) activities that can build and strengthen the community of computer science educators in India. Many ideas were generated, and iSIGCSE plans to adopt them and use them to guide activities throughout the year.

\end{abstract}

\section{Introduction}
\label{sec:introduction}

The focus of COMPUTE is to improve the quality of computing education in India by providing a platform for academicians and researchers to interact and share best practices in teaching, learning, and education in general. COMPUTE 2025 (held at IIT Ropar, 7-9 December 2025) introduced a Best Practices track to create an opportunity to share teaching best practices in computer science at the higher education level, in a setting that is less formal than a paper presentation. The goal of this track is to invite CS educators to share practices they have used in their classrooms and to disseminate them so that other educators can reuse or adapt them. It is also expected that this will lead to collaborations that extend these practices and experiences into research initiatives, and help ease the transition from being an educator to being a researcher.

In the first year of the track, we received sixteen submissions. Of these, three were accepted as Best Practices papers and two were accepted as ideas for further discussion. All authors were invited to present in Best Practices sessions at COMPUTE 2025.

The Best Practices papers have been published:
\begin{itemize}
\item Functional Python Programming in Introductory Computer Science Courses \cite{sunderraman2025functionalpythonprogrammingintroductory}
\item SocraticAI: Transforming LLMs into Guided CS Tutors Through Scaffolded Interaction \cite{sunil2025socraticaitransformingllmsguided}
\item Ancient Algorithms for Modern Curriculum \cite{thakkar2025ancientalgorithmsmoderncurriculum}
\end{itemize}

The Best Practices ideas were discussed in a session organized by the Best Practices chairs, with the goal of gathering more ideas to improve teaching and learning practices and to strengthen the CS educators community in India.

Section \ref{sec:ideas-session} summarizes the ideas session, and Appendix \ref{ideas-list} lists the detailed ideas shared by the attendees. These ideas are expected to be adopted by ACM iSIGCSE (the India chapter of the Special Interest Group on Computer Science Education) in its activities throughout the year, and the outcomes will be presented at the next COMPUTE conference.

\section{Ideas Session}
\label{sec:ideas-session}

We ran a session in which we identified three areas (listed below) and invited ideas from the community members attending the session. We selected two areas related to programming because it is one of the most commonly discussed courses and is of interest to all attendees. The third area focused on building the CS education community.

\begin{enumerate}
\item Area 1: Suggest class activities that can foster engagement and learning in programming classes.
\item Area 2: Suggest designs for take-home programming assignments that cannot be easily solved using AI agents, and are therefore largely AI-resistant.
\item Area 3: Suggest activities that can build and strengthen the community of computer science educators in India.
\end{enumerate}

More than fifteen ideas were shared in each area, which was very encouraging. We collated these ideas and shared them with the iSIGCSE core team, which will design year-long activities to help educators adopt and implement these ideas in their classrooms. The iSIGCSE core team also agreed to adopt the community-related ideas and include them in its activities. The following sections summarize the ideas we received in each area. Appendix \ref{ideas-list} contains the full list of ideas.

\subsection{Area 1: Class Activities}
A key message was to move from theory-heavy teaching to practice-oriented, engagement-driven learning. Participants strongly supported a shift toward hands-on work such as pair programming, debugging activities, unplugged exercises, gamification, project-based learning, and real-world problem solving. One attendee suggested a balance of roughly 80\% practice and 20\% theory, with frequent student-teacher interaction. Overall, the classroom was described as an active space for experimentation and collaboration, rather than passive content delivery.

\subsection{Area 2: Take-home assignments (that are AI-resistant)}
A key message was to redesign assessment so that learning is validated through process, reasoning, and interaction, not only through final answers. Participants questioned traditional take-home assignments and suggested alternatives such as viva voce, live coding, peer review, storytelling-based problems, modular team tasks, and AI-resilient assessments. Rather than banning AI, the discussion leaned toward controlled and transparent use of AI, combined with explanation and critique of solutions. The overall view was that assessment should measure understanding and creativity, not just output correctness.

\subsection{Area 3: Community}
A key message was that strengthening CS education requires strong, active communities of educators rather than isolated efforts. The discussion emphasized creating regional champions, regular meet-ups, workshops, and online platforms where teachers can share teaching resources, classroom activities, and best practices. Teacher networks, FDPs (Faculty Development Programs), ACM chapters, and community portals were seen as important for ongoing professional growth. The underlying view was that sustained improvement in CS education will depend on collaborative educator communities.

\section{Conclusion}
\label{sec:conclusion}

In 2025, the COMPUTE conference introduced a Best Practices track to strengthen the computer science education community in India by providing a forum to share and discuss best practices in teaching and learning. This paper summarizes the track activities and lists the ideas shared by community members. iSIGCSE will adopt these ideas and support them through activities held throughout the year. Feedback from the community has been very positive. Participants appreciated the high level of interaction and engagement in the Best Practices sessions, and we plan to expand the track further in future editions of COMPUTE.

\appendix
\section{Ideas shared by the community}
\label{ideas-list}

\subsection{Class activities}
\begin{figure}
    \centering
    \includegraphics[width=0.9\linewidth]{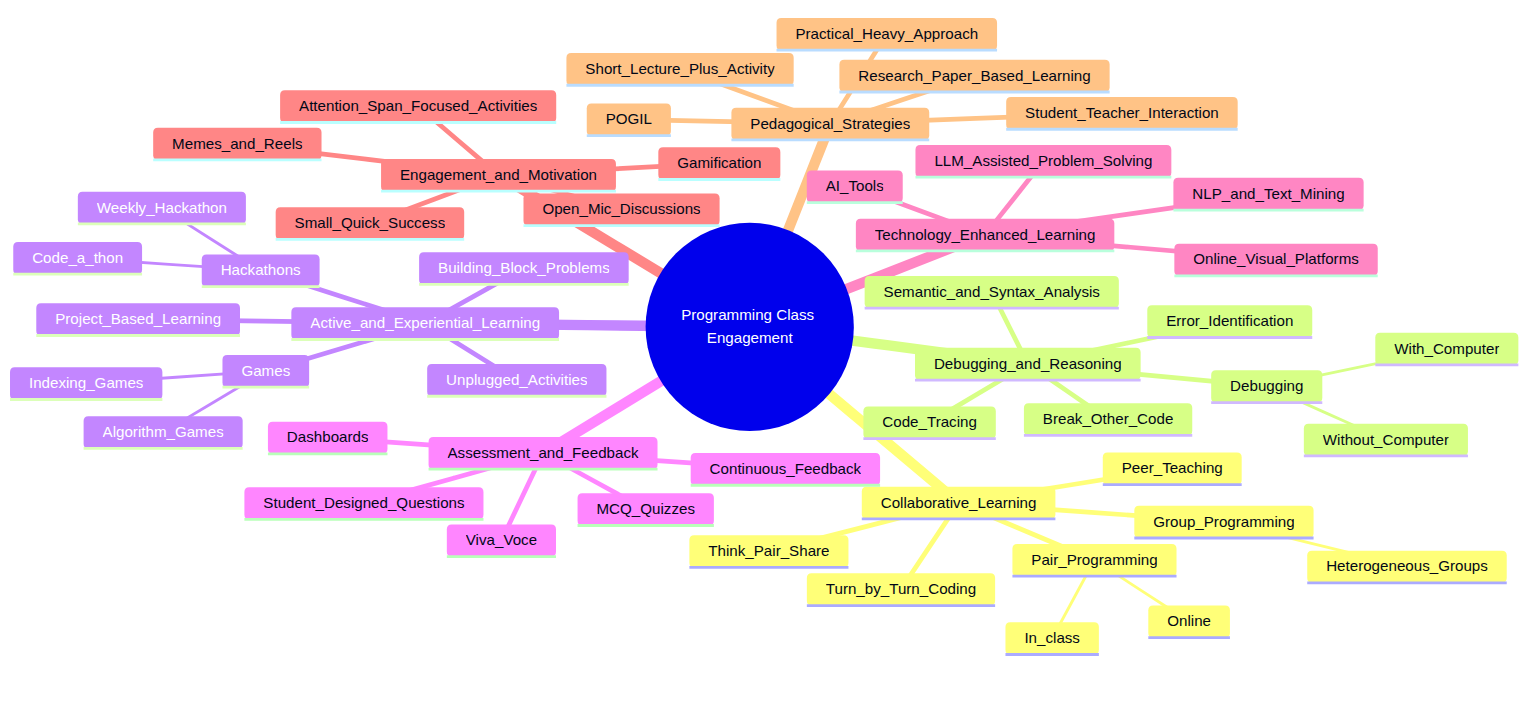}
    \caption{Suggested Class Activities}
    \label{fig:class-activities}
\end{figure}

Participants emphasized the importance of active, collaborative learning in programming education, advocating a shift away from lecture-centric instruction toward student-centered activities that promote reasoning, discussion, and problem solving (fig. \ref{fig:class-activities}). Pair and group-based programming, particularly for debugging, emerged as a core strategy. Activities such as pair programming, code tracing, Think–Pair–Share, peer explanation of errors, and solution critique were viewed as effective for developing both syntactic and semantic understanding of code. The use of activity-based and project-based learning was strongly supported. Participants suggested project work, heterogeneous group programming, small hackathons, and algorithm-focused games to facilitate deeper conceptual understanding and teamwork.

To sustain engagement, gamification and interactive tools were recommended, including creative tasks, AI-assisted problem solving, online visual platforms, and dashboards. Given students’ limited attention spans, frequent interaction and immediate feedback were considered essential. Assessments were framed as formative and continuous, with recommendations for low-stakes quizzes, student-generated questions, pre-class group assignments, and individual oral assessments. Overall, participants advocated for a practice-oriented approach, emphasizing minimal theory, extensive hands-on work, strong student–teacher interaction, and occasional unplugged activities for conceptual engagement.

\subsection{Take-home assignments (that are AI-proof)}

\begin{figure}
    \centering
    \includegraphics[width=0.8\linewidth]{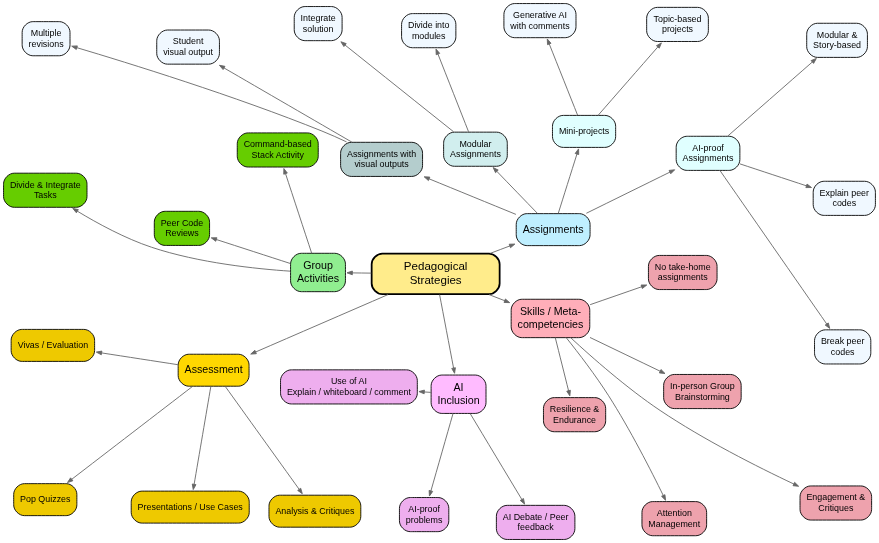}
    \caption{AI-Proof Assignments}
    \label{fig:assignments}
\end{figure}

Participants were largely skeptic toward traditional take-home assignments, noting concerns related to AI misuse, attention span, and authenticity of learning (fig. \ref{fig:assignments}). Several attendees suggested minimizing or eliminating take-home work in favor of in-person assessments such as group brainstorming, viva voce examinations, and endurance-based tasks that better reveal student understanding and resilience.

Where take-home assignments are used, most of the members emphasized process-oriented and transparent AI use. Students could be required to submit AI interaction logs, with evaluation focusing on the quality of prompts, critiques, and reflective engagement rather than the final output alone. Assignments should require students to analyze program design decisions and explain their reasoning through presentations or whiteboard discussions.

To increase resistance to AI-generated solutions, participants proposed contextualized, personalized, and visual tasks. Examples included programs analyzing a student’s own recent activities, assignments requiring complex visualizations, story-embedded problems, and the use of in-house tools unfamiliar to public AI systems. Such tasks were viewed as difficult to solve without multiple revisions and human judgment.

Collaborative and oral assessment strategies were also strongly supported. Suggested approaches included team projects with individual responsibility for submodules, peer teaching, peer code critique, and in-class activities where students identify, refute, or break others’ code. Viva voce examinations were recommended as a final verification step, particularly for testing, constraints, and expected outcomes.

Finally, participants supported guided and critical use of AI rather than outright prohibition. Students could be allowed to use generative AI but required to add innovative elements, justify decisions, debate whether solutions are AI-generated, or assess peer work for authenticity. Overall, AI-resistant assignments were characterized by modular design, personalization, oral explanation, peer interaction, and an emphasis on reasoning over mere code submission.

\subsection{Community Ideas}

\begin{figure}
    \centering
    \includegraphics[width=1.0\linewidth]{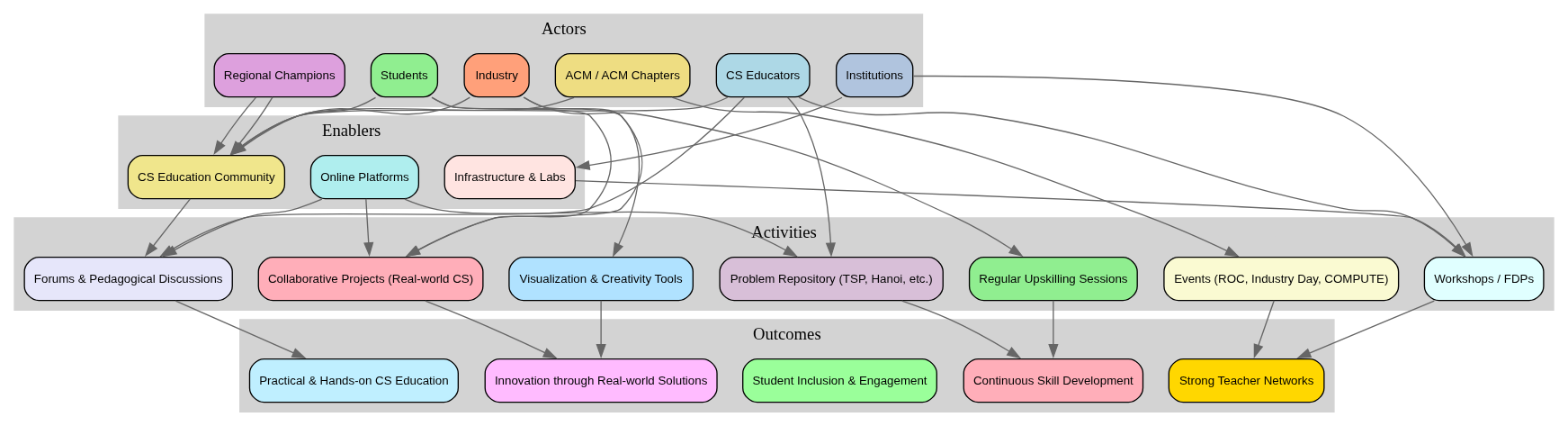}
    \caption{The Community Ideas Ecosystem}
    \label{fig:community-ideas}
\end{figure}

To develop the community further (fig. \ref{fig:community-ideas}), the participants emphasized the need to strengthen regional and national communities of practice for computer science educators in India. Regular, small-scale events, such as regional meetups, industry interaction days, and practice-sharing sessions, were recommended across zones. Participants proposed developing a dedicated community platform for CS educators to share pedagogical tools, classroom activities, and problem sets, while also enabling ongoing conversations and both formal and informal networking. Towards the same, identifying regional champions to organize in-person activities was viewed as critical for sharing classroom techniques, contextual best practices, and locally relevant pedagogical innovations.

Knowledge sharing through structured observation and documentation was another key theme. Classroom exchanges and a centralized repository or website were suggested to capture and disseminate successful pedagogical experiments, along with a curated collection of canonical CS problems (e.g., Tower of Hanoi, Traveling Salesperson Problem) and their instructional variations.

Professional development was framed as continuous and practice-oriented. Participants proposed multi-day workshops with a structured format combining conceptual grounding, hands-on implementation, student-led presentations, and feedback. They also recommended collaborative online forums where educators jointly work on real-world CS problems, benefiting from diverse perspectives and expertise.

Finally, participants highlighted the importance of sustained upskilling and institutional support. Strong teacher networks, faculty development programs, local ACM chapters, and online forums were identified as enablers of long-term capacity building. Given the rapidly evolving nature of CS, regular upskilling, practical-first curricula supported by adequate infrastructure, and the use of creative visualization tools for teaching algorithms were seen as essential to strengthening the CS education ecosystem.

\bibliographystyle{splncs04}
\bibliography{practices}

\end{document}